\documentclass[conference]{IEEEtran}
\usepackage{graphicx}
\usepackage{times} 
\usepackage{mathptmx}

\begin{document}

\setlength{\parskip}{1em}

\title{Interoperability and Standardization of Intercloud Cloud Computing}

\author{\IEEEauthorblockN{Jingxin K. Wang}
\IEEEauthorblockA{Dept. of Software Engineering\\
Harbin Institute of Technology\\
Weihai, PR. China, 264209\\
Email: kinghsin.wang@ieee.org}
\and
\IEEEauthorblockN{Jianrui Ding}
\IEEEauthorblockA{Dept. of Software Engineering\\
Harbin Institute of Technology\\
Weihai, PR. China, 264209\\
Email: dingjianrui@163.com}
\and
\IEEEauthorblockN{Tian Niu}
\IEEEauthorblockA{School of Business and Management\\
Hong Kong Univ. of Sci. \& Tech.\\
Kowloon, Hong Kong SAR\\
Email: sophia0203nt@gmail.com}
}

\maketitle

\begin{abstract}
Cloud computing is getting mature, and the interoperability and standardization of the clouds is still waiting to be solved. This paper discussed the interoperability among clouds about message transmission, data transmission and virtual machine transfer. Starting from IEEE Pioneering Cloud Computing Initiative, this paper discussed about standardization of the cloud computing, especially intercloud cloud computing. This paper also discussed the standardization from the market-oriented view.
\end{abstract}


\IEEEpeerreviewmaketitle
\section{Introduction}
\subsection{Intercloud Cloud Computing}
With cloud computing applications being used more and more widely, there becomes more and more cloud computing service providers. However, whether the cloud computing services provided by different providers can interoperate, whether them have a common interface, has become a problem to be solved.

The Intercloud is an interconnected global "cloud of clouds".\cite{interoperability} It can provide an extension of computing and storage capacity to a single cloud.
\begin{figure}[h]
\centering
\fbox{\includegraphics[width=2.5in]{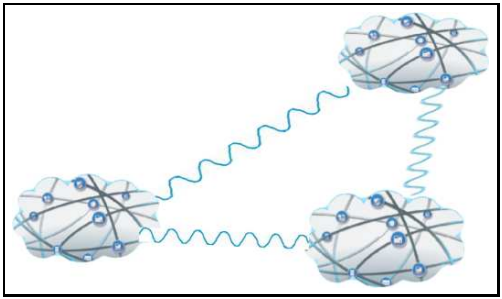}}
\caption{Intercloud Cloud Computing Model}
\label{fig_01}
\end{figure}

\subsection{Interoperability and Standardization}
\subsubsection{Interoperability}
Imagine the following scenarios: a company using a cloud application despoilment in cloud computing provider A, then the servers of provider A crashed, the company should transfer the application to provider B. Here the cloud computing service provided by A and B need guarantee the interoperability to each other. 

The interoperability among clouds should include message transmission, data transmission, virtual machine transfer, etc..

\subsubsection{Standardization}
In order to guarantee the interoperability among different cloud computing platforms, it is essential to work out standards to describe the cloud itself, the interface to communicate, and the data format. There are already several standardization organizations working on the cloud computing standards, such as IEEE Standard Association, ISO/IEC JTC1 SC38, etc..

\section{Interoperability of Clouds}
\subsection{Message Transmission}
Applications which run on several different cloud computing platforms may send, receive, and process messages from each other from time to time. The message transmission is an important factor to guarantee program runs regularly. Just like the kind of Instant Message software (IM software), cloud applications also need the ability to communicate with each other by sending message.

Differently from the common IM software, the cloud applications may communicate with each other via different protocols. The protocol best satisfies the demands here is XMPP, which is defined by the Internet Engineering Task Force (IETF) in RFC6120: "Extensible Messaging and Presence Protocol (XMPP): Core"\cite{RFC6120} , RFC6121: "Extensible Messaging and Presence Protocol (XMPP): Instant Messaging and Presence"\cite{RFC6121}, RFC6122: "Extensible Messaging and Presence Protocol (XMPP): Address Format".\cite{RFC6122}

\begin{figure}[h]
\centering
\fbox{\includegraphics[width=2.5in]{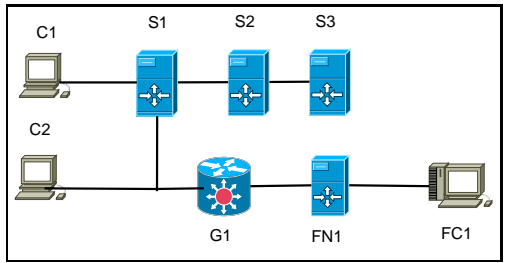}}
\caption{Interaction Between XMPP and Other Protocol\cite{RFC6120} }
\label{fig_02}
\end{figure}

In figure 02, C1, C2 are XMPP clients, S1, S2  are XMPP servers, G1 is a gateway that translates between XMPP and the protocol(s) used on a foreign (non-XMPP) messaging network, FN1 is a foreign messaging network, FC1 is a client on a foreign messaging network.\cite{RFC6120} 

This protocol could be used to communicate with different cloud computing platform, meanwhile, XMPP can even be the control protocol to administrate the cloud resources.

\subsection{Data Transmission}
Cloud computing provide us the capacity to process mass data,and this also brings some issues about transmitting the mass data. The cloud computing provider can also provide cloud storage service, providers offer many datacenter to store users' data. In the model with cloud exchange root\cite{intercloud_root} , cloud service providers offer the interface for users to communicate with cloud exchange root, so the private clouds owned by users could exchange data with root servers.

\begin{figure}[h]
\centering
\fbox{\includegraphics[width=2.5in]{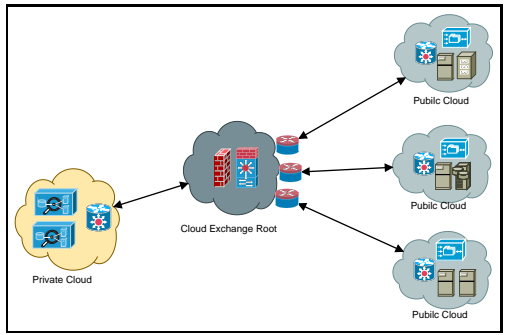}}
\caption{Topology of Data Transmission in Cloud Exchange Root Model}
\label{fig_03}
\end{figure}

The cloud exchange root servers can balance loadings and requests for the storage engine. This can provide a better performance to users and easier to extend for providers. Yet like the scenario shown in figure 02, providers may store the data into different platform and format, and this causes difficulties for users process them.

A kind of platform independent and operating system independent data format is needed here to exchange among different storage solution engines. This kind of format is called Uniform Data Format (UDF). UDF is not a kind format to describe concrete data like JPEG or mp3 formats do, it is like a kind of archive file to describe attribution of the files and also can package the file for transmission.

\subsection{Virtual Machine Transfer}
Some PaaS providers offer the platform to users to build their own virtual machine. As for these PaaS providers, an important aspect of interoperability is capacity of transferring the virtual machine.

\subsubsection{Complete Virtual Machine Platform}
Some platforms allow users to establish their own virtual machine and install the complete operating system, these platforms are called complete virtual machine platform.

Virtual machine transfer on the complete virtual machine platform should satisfied the virtual disk file of the virtual machine is common format and the cloud platforms have same Hardware Abstract Layer (HAL). And the configure file of the machine should be self-adapting when the disk was transfered to another cloud platform.

\begin{figure}[h]
\centering
\fbox{\includegraphics[width=2.5in]{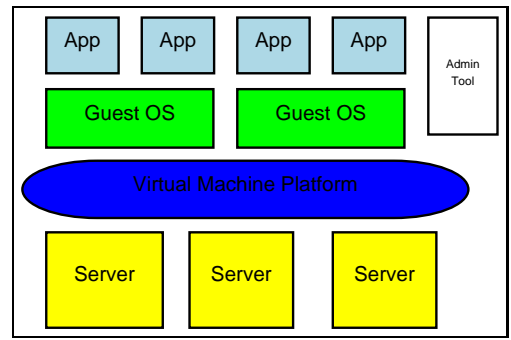}}
\caption{Complete Virtual Machine Platform}
\label{fig_04}
\end{figure}

\subsubsection{Run-time Environment}
Some platforms do not provide the virtual machine platform to allow users install their own operating system, they provide a set of API instead. The platform offers the HAL to isolate the difference of the hardware, and build a gateway to allow the cloud communicate with others.

This model need the uniform interface to allow the applications run on different platform. The run-time environment can provide the uniform interface, like the Java virtual machine, the run-time environment and the HAL can provide the platform-independent API for SaaS companies and developers.

However, as for run-time environment model, there are still some difficulties. First, there is a gateway in the model, and it is above the HAL, so different run-time environment may have different gateway, is can limit the universality of the applications. And platform-independent API may meet the problems of compatibility.

\begin{figure}[h]
\centering
\fbox{\includegraphics[width=2.5in]{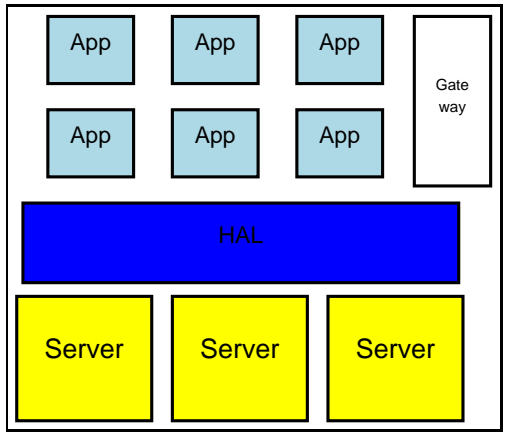}}
\caption{Run-time Environment}
\label{fig_05}
\end{figure}

\section{Standardization of Cloud Computing}
\subsection{IEEE Pioneering Cloud Computing Initiative}
The Institute for Electrical and Electronics Engineers (IEEE), announced the new initiative to serve as the catalyst for innovation in the cloud computing arena in 2011\cite{IEEE} , which included two new standards development projects, IEEE P2301™, Draft Guide for Cloud Portability and Interoperability Profiles, and IEEE P2302™, Draft Standard for Intercloud Interoperability and Federation.

IEEE P2301 will aid users in procuring, developing, building, and using standards-based cloud computing products and services, enabling better portability, increased commonality, and greater interoperability across the industry and provide an intuitive road map for cloud vendors, service providers, and other key stakeholders. Meanwhile, IEEE P2302 defines essential topology, protocols, functionality, and governance required for reliable cloud-to-cloud interoperability and federation, which will help build an economy of scale among cloud product and service providers that remains transparent to users and applications.\cite{IEEE} \cite{P2302}

Meanwhile, other standard organizations also launched some cloud computing standardization projects, such as ISO/IEC JTC1 SC38 Distributed application platforms and services (DAPS), it established Cloud computing working group to work on the cloud computing standards.

\subsection{Market-Oriented View of Standardization}
\subsubsection{De Facto Standards}
The commercialization of the cloud computing happened earlier and distributed. Several companies are already providing cloud computing services. Like Amazon, it has Amazon Web Services (AWS), which offers a complete set of infrastructure and application services that enable consumer to run virtually everything in the cloud.\cite{AWS} Meanwhile, Microsoft and Google also have their cloud services, like Windows Azure and Google App Engine.\cite{WA} \cite{GAE} These companies and their solutions are influential in cloud services market, the regulation their used becomes de facto standards.

\subsubsection{Accurate Description of Cloud Services}
Another important mission of standardization of cloud computing is providing a set of description of cloud services to the cloud consumers, cloud creators, and cloud providers. Like IEEE P2301 does, this kind of standard should define the common description to let cloud consumers be aware of what kind of services they can get accurately. This can standardize norms for the cloud services' market.

\subsubsection{Cloud Computing Ecosystem}
Not only the cloud infrastructure providers can offer complete cloud services. Cloud infrastructure providers have to offer the open platform and APIs to developers to allow them develop applications on their platform. Just like the mobile application store, such as Apple App Store, the providers should supply such kind of platform and SDK for application developers.

The important aspect of promoting the healthy development of the cloud computing market is to establish a upstanding cloud computing ecosystem. This requires IaaS, PaaS and SaaS providers a closer links. Like the PC ecosystem, now the IaaS providers like the hardware manufacturers, PaaS providers like the operating system companies, and SaaS providers like the application companies and developers. 

Meanwhile, in order to guarantee essential interoperability among clouds, the cloud platform should under the same standard of the user interface even APIs. Unlike the operating system, cloud consumers may change the services provider especially the original one crashed temporarily. The intercloud cloud model is designed to solve these kind of problems. Intercloud model can supply cloud of clouds, which allow users transfer their services to a cloud which operates normally when the original one stops services. Because intercloud model often offered inter-company, this would satisfy the interoperability of the platforms.

\subsubsection{Interoperability and Security}
Cloud computing network is a kind of open network, which is faced with various security threats. Under thee requirement of interoperability, the cloud network need to ensure the security from aspects of encryption of communication protocols and trust model based on Public Key Infrastructure (PKI) and authentication.

As for encryption of communication protocols, XMPP has already provided the encryption characteristic in its standard.\cite{RFC6120} XMPP supports the Simple Authentication and Security Layer (SASL) framework and Transport Layer Security (TLS) protocol. The SASL framework provides a protocol for securing subsequent protocol exchanges within a data security layer.\cite{RFC4422} The TLS protocol allows client/server applications to communicate in a way that is designed to prevent eavesdropping, tampering, or message forgery.\cite{RFC5246}

The Intercloud Root in intercloud cloud computing model can serve as a Trust Authority. \cite{intercloud_economy}

\begin{figure}[h]
\centering
\fbox{\includegraphics[width=2.5in]{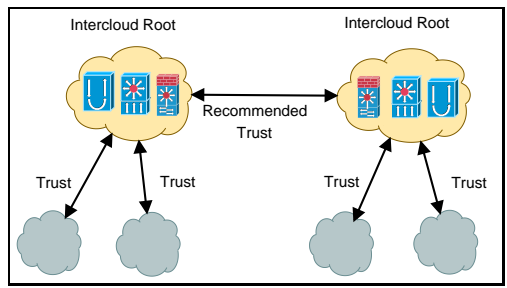}}
\caption{Intercloud Trust Management Model}
\label{fig_06}
\end{figure}

As figure 06, the clouds under the same Intercloud Root can trust the Root directly, the Root maintains the trust list as a Certificate Authority (CA), and can issue certificate to clouds in its domain. Between two Roots, they can recommend each other to trust the other. When a cloud is added in the domain, the Intercloud Root should check and authenticate the cloud, if it is trustable, the Root will issue a certificate for it. When cloud A apply the storage resources from cloud B, cloud A should send the request to the Root of the domain, if cloud A and B are under same domain, the Root would check the trust list and send the result to cloud A. Otherwise, if cloud B is under other domain, Root of domain A will send the request to the Root of domain B, Root B will send Root A its recommendation. Then Root A will send the recommendation to cloud A. Cloud A can choose the trust level of cloud B.

\section*{Acknowledgment}
The first author would like to thank the outstanding work of colleagues in IEEE Standard Association P2301 and P2302 working group. And authors also would like to thank the reviewers for their very helpful comments and suggestions.

\nocite{*}
\bibliographystyle{IEEEtran}
\bibliography{IEEEabrv,ISICC.bib}

\end{document}